# Generation of Web Apps with Agentic IDEs: An Empirical Assessment

Manuel Marceca
University of Milano-Bicocca
Milano, Italy
m.marceca@campus.unimib.it

Maria Teresa Rossi
Gran Sasso Science Institute
L'Aquila, Italy
mariateresa.rossi@gssi.it

Leonardo Mariani
University of Milano-Bicocca
Milano, Italy
leonardo.mariani@unimib.it

## Abstract

Agentic IDEs are among the most significant innovations in software engineering, aiming to accelerate application development through LLM-based agents that can assist developers during development. However, their evaluation in end-to-end development tasks involving the generation of complete applications remains limited. To fill this gap, we propose a rigorous comparative analysis of three popular agentic IDEs (Copilot, Cursor, and Windsurf) in the generation of five full-stack Web applications from scratch.

Results show high maturity in the generation of established patterns, such as CRUD operations and authentication features. In contrast, the generation of less common distributed architectures, such as a task queue architecture, produces significantly more errors. Overall, results show that Agentic IDEs cannot replace developers but shift their role toward building software by orchestrating LLM-based agents through natural-language instructions and iterative refinement. Yet, each agentic IDE shows its peculiarities, although differences are narrow.

## 1 Introduction

Recent advances in large language models (LLMs) have rapidly transformed AI-assisted software development from syntax-level autocompletion [32] into interactive, tool-augmented systems that can plan large-scale changes, modify multiple files, execute commands, and iteratively refine solutions inside modern development environments [22]. Surveys and industrial reports indicate that such assistants are already integrated into professional workflows, affecting activities ranging from feature implementation to testing and documentation [33, 35]. This evolution has stimulated the emergence of *agentic IDEs*, i.e., development environments in which the assistant is not limited to proposing local snippets but can act semi-autonomously across the project context, with the developer supervising, correcting, and validating results [37]. Indeed, agentic IDEs are changing the locus of engineering effort, which is gradually moving from code implementation to verification, debugging, integration, and code auditing (e.g., for security [17]).

Despite the rising interest, the practical capabilities of agentic IDEs remain insufficiently characterized in development scenarios that resemble end-to-end application engineering. For instance, several studies focused on LLM-based code generation, revealing complementarities between models, yet focusing on small-scale code generation tasks [30, 38, 44]. Benchmark suites for code synthesis and repair, such as HumanEval [14] and SWE-bench [27], provide important standardized testbeds, yet they typically emphasize isolated programming tasks or issue-resolution settings that do not fully capture the interactive, multi-component, full-stack nature of application construction. Some studies considered user perceptions and adoption patterns [15, 33, 35], instead of the correctness of the generated code. In a nutshell, evidence on whether agentic IDEs can reliably generate complete full-stack applications, including, backend logic, persistence layer, security constraints, interactive frontend behavior, and automated testing, is still limited.

In this paper, we focus specifically on agentic IDEs because IDE integration represents one of the most practically relevant modalities through which developers currently interact with coding agents [1, 37]. In an agentic IDE, code editing, project navigation, command execution, test execution, error inspection, and corrective prompting occur within the same workspace shared by the developer and the agent. This setting is particularly relevant for supervised software generation, where the developer remains in the loop and iteratively validates and refines the generated artifacts. Moreover, focusing on IDE-integrated agents provides a comparable interaction setting across tools, while still capturing realistic developer-agent workflows. Accordingly, our results should be interpreted as evidence about IDE-integrated agentic workflows, rather than as claims about coding agents independently of the development modality in which they are used.

This paper investigates the capabilities of three well-known agentic IDEs (GitHub Copilot [4], Cursor [3], and Windsurf [8]) across five application scenarios of increasing architectural and operational complexity: a CRUD note manager, an authentication module, a secure file hosting module, a real-time chat module with WebSocket communication, and a distributed task-queue system. In this context, we study the overall effectiveness of the agentic IDEs (RQ1), their capability to iteratively improve the generated code (RQ2), the efficiency of the generation process and the human effort, needed to assess it (RQ3), and the impact of model evolution on the generation process (RQ4). To respond to these research questions, we analyze a total of 48 automatically-generated project implementations, which are carefully inspected both manually and automatically to respond to the research questions.

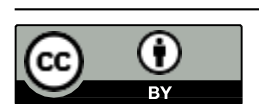

Results show that with established and common features, such as CRUD operations and authentication, agentic IDEs achieve high performance, with a low error rate, and few corrective iterations necessary to produce the code. In contrast, less established applications with more complex architectures remain challenging, with a higher error rate. A dominant and persistent limitation is automated test-suite generation, which accounts for 49.6% of all observed errors. Comparative analysis shows that differences between agentic IDEs are narrow, with trade-offs primarily concerning speed and correctness. Some IDEs such as Cursor tend to prioritize faster generation times, while others such as Windsurf adopt a more deliberate approach, producing more complete solutions that require fewer corrective iterations, highlighting a persistent gap between speed-oriented and correctness-oriented strategies. Performance also varies across layers, showing that test generation is one of the most significant challenges. Overall results show that the role of the developers is gradually shifting toward building software by orchestrating Agentic IDEs through natural-language instructions and iterative refinement.

In a nutshell, this paper provides the following key contribution:

- It presents an empirical study about the capabilities of three state-of-the-art Agentic IDEs;
- It shows how Agentic IDEs perform with five application of different complexity, also investigating the quality of the individual code layers that have been generated;
- It considers a range of metrics and perspectives that combine automatic measurements with human-judgment checks, which is a distinctive aspect of our study;
- It delivers a set of actionable findings and final takeaways that can influence future research in the area;
- It delivers the experimental material to support the reproduction and extension of the proposed study.

The paper is organized as follows. Section 2 positions our work with respect to the state of the art. Section 3 describes the investigated research questions and the methodology defined to address them. Section 4 presents empirical results. Sections 5 and 6 discuss how results influence research and practice in the area. Finally, Section 7 provides final remarks and discusses future work.

## 2 Related Work

***LLM-Based Tools*** The rapid evolution of artificial intelligence has transformed software engineering practices, especially through the emergence of *LLM-based* tools, which are capable of generating code, explaining implementations, and assisting developers during multiple phases of the software development lifecycle [14, 19, 29, 34]. As a result, AI-powered assistants have increasingly been integrated into development environments, giving rise to a new paradigm of AI-augmented software development.

LLM-based tools have known limitations. For instance, one well-documented issue concerns the generation of incorrect or incomplete code that appears plausible but fails in the real execution conditions [18]. In fact, empirical analyses have shown that LLM-based code generators may struggle when tasks involve long-range dependencies, multiple interacting components, or repository-level reasoning. Moreover, studies on AI-generated code have identified security risks and quality issues, showing that automatically generated code can contain vulnerabilities or unsafe programming patterns if not carefully reviewed [20, 39]. These findings emphasize the importance of investigating the quality of AI-generated code in a range of use cases.

LLMs have also been investigated in automated test generation tasks. Benchmarks such as TestBench [43] and subsequent approaches based on hybrid analysis or coverage-guided generation [25, 40] show that, although LLMs can produce syntactically valid tests, achieving high coverage and correctly handling edge cases, asynchronous behavior, and I/O-intensive scenarios remains challenging. These findings are consistent with our results, which identify test generation as one of the main limitations of current agentic IDEs.

***Modern IDEs and Their Assessment*** *Modern AI-assisted development environments* typically embed neural code generation and contextual completion mechanisms directly within the IDEs. Systems such as GitHub Copilot [4], CodeWhisperer [2], and ChatGPT-based assistants [6] provide real-time suggestions based on the developer's code context and textual prompts. Empirical studies indicate that these tools can significantly improve productivity, particularly for repetitive tasks, boilerplate generation, and documentation writing [31]. At the same time, user studies investigating developers' interactions with such systems show that programmers increasingly treat AI assistants as collaborative partners that support ideation, debugging, and exploration during development workflows [38]. These findings suggest that AI-assisted programming is progressively evolving from simple autocomplete functionality toward a form of human–AI pair programming.

To assess the performance of these tools, several *benchmarks* and evaluation frameworks have been defined. Widely adopted benchmarks include HumanEval [14], which can be used to evaluate the functional correctness of AI-generated code through unit tests, and MBPP [12], a dataset designed to assess program synthesis capabilities from natural language prompts. More recently, SWE-bench has introduced a realistic evaluation setting in which language models must resolve real issues extracted from open-source GitHub repositories, thereby reflecting more complex software engineering scenarios [27]. These benchmarks have played an important role in standardizing evaluation practices. However, none of these benchmarks encode *code-generation tasks at scale*, such as the generation of entire applications.

Beyond function-level benchmarks and issue-resolution settings, recent work has started to investigate repository-level and multi-file code generation. Approaches such as RepoCoder [42] and CodePlan [13] highlight the importance of leveraging cross-file context and planning to support coordinated edits, while recent benchmarks such as RepoExec [28] explicitly evaluate executability and correctness at the repository level. These studies move evaluation closer to realistic software engineering scenarios, yet they mainly focus on code completion or repository modification tasks rather than the generation of complete applications from scratch, as we do.

Results show that while LLMs demonstrate strong performance on localized programming tasks, their effectiveness decreases when they must coordinate across several modules, manage stateful interactions, or maintain architectural consistency across large codebases. Recent research evaluating AI-assisted development tools highlights variability in the quality of generated code depending on prompt formulation, model configuration, contextual information, and the structure of the development task itself [9, 21, 24]. These challenges indicate that, although the underlying models are rapidly improving, fully autonomous software development remains an open research problem.

***LLM-Based Agents*** A novel research direction is *agent-based software engineering*, where language models are embedded in autonomous or semi-autonomous agents capable of interacting with development tools, navigating repositories, executing tests, and iteratively refining code [26]. Frameworks such as ReAct integrate reasoning and action to enable LLMs to plan multi-step problem-solving strategies [41]. Systems like SWE-agent extend this paradigm by providing specialized interfaces that allow language model agents to operate directly within software development environments [27]. More broadly, recent research on LLM-based multi-agent systems for software engineering [26] frames these approaches as a shift from passive code suggestion toward autonomous or semi-autonomous workflows, demonstrating that LLM agents can potentially address complex software engineering tasks by combining code generation with planning, environment interaction, and iterative refinement.

Iterative refinement and self-correction are also extensively investigated. Techniques such as Reflexion [36] and related self-debugging approaches [16] show that agents can improve performance by incorporating feedback across multiple attempts, highlighting the importance of iterative loops in achieving correct solutions. While these mechanisms are increasingly studied in controlled benchmarks, their impact on the generation of complex, multi-layer software systems remains under-explored. In our empirical study, we investigate iterative refinement of the solution by letting the agents improve the generated implementation.

***Generation of Full Stack Applications with Agentic IDEs*** Within this evolving landscape, a key research gap concerns the systematic evaluation of *agentic IDEs* [3, 4, 8] that integrate LLMs with automated agentic workflows controlled from the IDEs to generate, modify, and test software applications. While existing benchmarks and empirical studies provide valuable insights into isolated tasks or individual tools, the performance of such systems in end-to-end software development scenarios involving complete application modules, multiple architectural layers, and iterative correction processes is under-explored. Consequently, understanding the strengths and limitations of these tools in realistic large-scale development scenarios remains an important open problem.

In this context, our study provides a controlled empirical evaluation of *agentic IDEs in the end-to-end generation of complete full-stack Web applications, explicitly analyzing multiple architectural layers (frontend, backend, persistence, and tests) together with iterative correction behavior and human validation effort.* By analyzing performance across multiple complexity levels, the study generates insights on the capabilities and limitations of Agentic IDEs.

# 3 Methodology

## 3.1 Research Questions

Agentic IDEs combine LLMs with an execution layer that can navigate a codebase, create and modify multi-file projects, run commands, interpret runtime output, and iteratively refine solutions. While this paradigm promises to compress the end-to-end software lifecycle into a tightly coupled human–agent workflow, its practical effectiveness depends on whether these agents can consistently satisfy functional requirements, maintain architectural coherence across iterations, and avoid introducing regressions while fixing defects. The goal is thus to investigate to *what extent agentic IDEs can be used to address the challenging task of generating a full stack application*. This goal is mapped into four research questions.

**RQ1 (effectiveness): How *effective* are agentic IDEs at generating Python/JavaScript full-stack Web applications?** This research question investigates whether agentic IDEs can produce a running application that satisfies the stated requirements and whose implementation is of high-quality.

**RQ2 (iterations): To what extent can Agentic IDEs *detect errors and improve* the generated code?** This research question investigates the iterative behavior of the agentic IDEs by considering the number of detected errors and the corrective actions taken to fix them, as well as the regressions that are introduced in the attempt to improve the code.

**RQ3 (efficiency): How *quickly* can the desired application be generated?** This research question investigates both the efficiency of the agentic systems, as well as the amount of human effort necessary to compensate for the inadequacies in the generated code.

**RQ4 (evolution): To what extent do results depend on the *evolution* of the underlying models?** The experimental campaign spans April 2025 to January 2026 and uses two successive generations of Claude Sonnet models in response to ecosystem availability. By repeating part of the experiments after eight months under updated models and IDE versions, our work aims to identify improvements attributable to model evolution from those due to IDE-specific engineering updates. In particular, we repeat twice the experiment with the Note Manager using Claude Sonnet 3.5 and Claude Sonnet 4.5 as underlying LLM, to appreciate the impact of model evolution on the results.

## 3.2 Agentic IDEs

We selected three Agentic IDEs that implement different integration patterns with IDEs and are frequently recognized as the best in the field [1]: the GitHub Copilot plugin-based agentic system (i.e., Copilot can be integrated into existing non-natively-agentic IDEs by installing its plugin), and the Cursor and Windsurf native Agentic IDEs (i.e., the IDE is natively implemented as an agentic system).

We describe below the three considered Agentic IDEs.

**Github Copilot [4]** is an AI-powered coding assistant supporting code generation from natural language, contextual autocompletion, and more recently agentic capabilities such as executing commands and iterating over solutions.

**Cursor [3]** is an AI-native IDE designed to tightly integrate agentic functionalities into the development environment. It provides advanced features such as autonomous agents that can navigate the codebase, execute terminal commands, and iteratively refine code starting from a single prompt.

**Windsurf [8]** is an AI-enhanced IDE featuring advanced agentic capabilities such as the Cascade agent for contextual assistance and multi-step task execution. It emphasizes a more structured and reflective interaction with the agent, supporting planning and coordinated execution across complex codebases, and has seen rapid industrial adoption.

We configured all the Agentic IDEs to run with Claude Sonnet 3.5, which is advertised as one of the best models for AI coding [10], to compare the capability of the agentic systems while using the same underlying model. We also use Claude Sonnet 4.5 [11], which has been released after we initiated our experiments, to address RQ4.

## 3.3 Experimental Subjects

The Agentic IDEs are used to generate applications from scratch. To avoid the data leakage problem, we consider the generation of entirely new applications, targeting the use case of developers who need to rapidly obtain a working system, not explicitly referring to any existing system. On one hand, we have intentionally written high-level functional specifications for the software to be implemented, so that the agentic IDEs are free to decide how to implement the system. On the other hand, we included strict boundaries on API endpoints (base paths, HTTP methods, required parameters), entity schemas (the use of certain fields), and testing requirements (including minimum test counts). Specifically, each generated project is required to include an automated test suite implemented using `pytest`, covering the core functionalities of the application, such as API endpoint behavior, data persistence, and main business logic. In some use cases, additional constraints are introduced, such as explicitly testing certain edge conditions. All the experimented agentic IDES receive the same instructions. Full prompts, including the experimental material to replicate our study, are available online[1].

We specifically identified five applications to be implemented, corresponding to increasing levels of architectural and operational complexity while remaining representative of realistic web development requirements. The selected applications are full-stack Web applications ranging from a simple data-management application, to authentication and file handling, up to real-time communication and asynchronous background processing. This design allows us to study how agentic IDEs behave when moving from consolidated request-response patterns to applications involving concurrency, state management, external storage, and coordination logic. In particular, we considered the following applications.

**Note Manager Application:** a baseline data-driven application implementing CRUD (create, read, update, delete) operations of a note entity, including a frontend interface and a real-time search/filter capability. The generation of this application is repeated twice in April 2025 and January 2026 to appreciate the impact of progress over models and IDEs. *Key characteristics:* Standalone application implementing the well-known domain of CRUD operations for a simple note entity.

[1] https://zenodo.org/records/19208143

**Authentication Module:** a complete authentication module with user registration, login/logout, profile retrieval, endpoint protection, and explicit security requirements such as password hashing and token-based authentication. *Key characteristics:* Standalone application implementing a well-known but technical security-related domain.

**File Hosting Application:** secure file upload/download/ management, including MIME validation, filename sanitization, storage handling, and prevention of filesystem attacks (e.g., path traversal), with a frontend supporting interactions such as drag-and-drop. *Key characteristics:* Standalone application that implements a relatively well-known domain and that has to interact with its environment to store and retrieve files.

**Real-time Chat:** a WebSocket-based chat system with concurrent connections, message broadcasting, persistence of rooms and messages, and a frontend supporting real-time interaction. *Key characteristics:* A distributed application that has to implement specific technical requirements about interaction with the network.

**Task Queue:** a distributed-style background processing system with asynchronous workers, task state management, retry logic, cancellation, monitoring, and real-time frontend progress visualization. *Key characteristics:* A distributed application that has to implement specific technical requirements about queues.

The generation of each application is repeated three times to address the stochasticity of the software generation process. This yields to 48 generated project instances overall: 5 projects × 3 Agentic IDEs × 3 repetitions plus 3 runs repeated for the Note Manager application with an updated model. To respond to the research questions, each generated project is systematically scrutinized both automatically and manually. This analysis corresponds to a significant amount of work, especially due to the many manual checks completed, that has been performed over several months of activity.

## 3.4 Experimental Design and Metrics

The experimental design is a controlled comparative study in which the primary independent variable is the *agentic IDE*, while the task specification, technology stack, validation protocol, and measurement procedures are controlled. An additional factor explicitly modeled in the analysis is the *functionality type*, which ranges from well-known domains (Note Manager application and Authentication Module) and relatively well-known domains (File Hosting application) to specific domains (Real-time chat and Task Queue).

*Protocol for Agentic Generation of the Applications.* Each project generation round follows a 4-stage iterative procedure that is shared among all the Agentic IDEs as shown in Figure 1.

**Stage① – Initial generation:** the IDE receives the full natural-language prompt and generates the complete multi-file application, including backend, frontend, persistence, and tests. Every prompt constrains the generated system to the same stack to ensure comparability across the projects generated by the agentic IDEs. This choice is motivated by the need to balance realism and experimental control. In particular, the selected technologies (FastAPI for the backend and plain HTML/CSS/JavaScript for the frontend) are widely adopted and well represented in LLM training data. The backend is implemented in FastAPI (Python) with REST endpoints and, when required, native WebSocket support. The frontend is implemented with plain HTML/CSS/JavaScript to avoid variability introduced by framework-specific toolchains; testing is performed with `pytest`. Persistence is required in all tasks and is realized via lightweight, portable mechanisms (e.g., SQLite or JSON-based storage), with task-specific storage extensions when required (e.g., filesystem storage for file hosting; additional components such as Redis may be used for queue-style coordination). Prompts also enforce a consistent repository layout with three top-level directories (`backend/`, `frontend/`, `tests/`), together with basic packaging artifacts (e.g., `requirements.txt`, README) to promote architectural separation and reproducible execution.

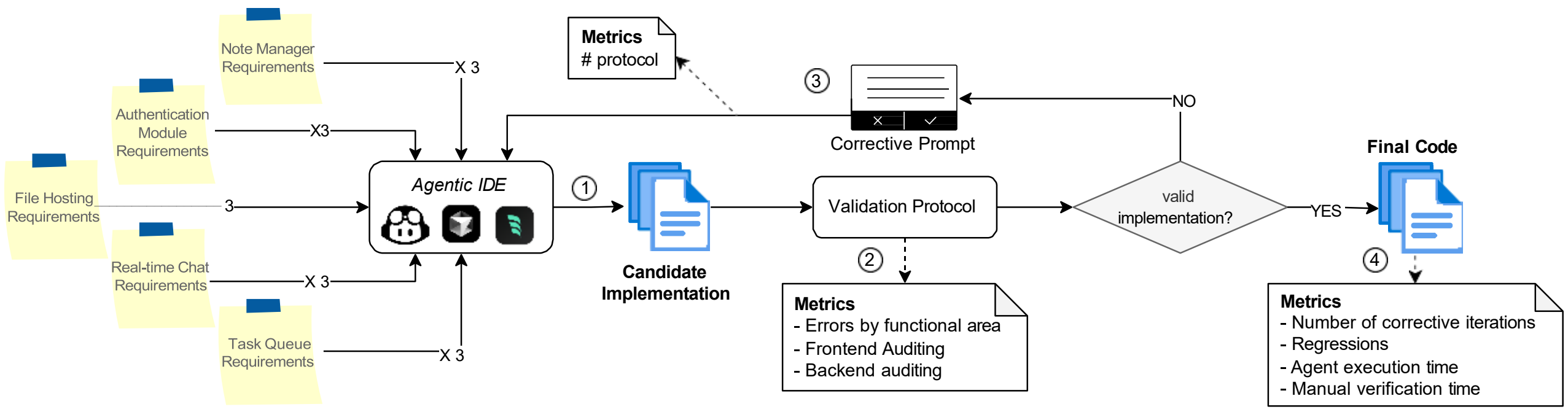

**Figure 1: Experimental Workflow.**

**Stage②– Validation:** the generated application is validated along six dimensions that cover the key elements: (i) installation, (ii) frontend interaction, (iii) backend API behavior, (iv) data persistence (v) presence and executability of automated tests, and (vi) project structure compliance. Validation combines deterministic automated checks and systematic manual inspection, as described in the validation protocol section.

**Stage③ – Correction:** if any requirement is violated during the validation process, a structured corrective prompt is provided to the agentic IDE. The prompt includes the observed deviations and the expected behavior. The prompt asks for only fixing the application with respect to the observed deviations, without introducing other changes. The agentic IDE produces an updated implementation that is validated again using the same protocol.

**Iteration:** the validation and correction states are repeated until all requirements are satisfied or for a maximum of 10 corrective iterations, at which point the generation process is considered non-convergent. The chosen cap guarantees termination. In our experiments, the generation always converged before the 10-round limit was reached.

**Stage④ – Data collection:** Throughout all phases, time measurements and outcome metrics are continuously recorded. Throughout all phases, we continuously record both time measurements and outcome metrics. Only interactions that involve actual corrective actions are counted as corrective iterations, while confirmation turns and IDE-internal review messages are excluded to ensure fairness across tools with different interaction styles.

*Validation Protocol.* The purpose of the validation protocol is to determine whether each generated project satisfies the requirements stated in the corresponding prompt and whether it can be executed as a complete full-stack application. The same protocol is applied to every generated project and to every corrective iteration. Validation combines executable checks, static analysis tools, and manual inspection, as follows:

*(1) Installation*. We install the declared dependencies and launch the backend and frontend components following the instructions and artifacts generated by the agent, such as `requirements.txt` and the project README. If the application cannot be installed or launched, the failure is recorded as a requirement-level violation and the observed problem is included in the corrective prompt.

*(2) Acceptance Testing*. We execute the frontend acceptance tests derived from the requirements. Since each generated application may have a different visual appearance, these tests are manually executed through the browser. Each acceptance test specifies the user interactions to be executed and the expected outcome. A failure is recorded if the frontend does not include the functionality under test, or the functionality produces a behavior that is inconsistent with the expectation.

*(3) Backend Testing*. We execute backend the API tests implemented according to API contracts. They validate endpoint availability, HTTP methods, input/output schemas, status codes, error handling, and the expected side effects on the application state. Failures detected by these scripts are recorded as backend/API requirement-level violations.

*(4) Persistence checking.* We check the data persistence layer by creating or modifying application data, stopping and restarting the application, and verifying whether the state is preserved as required.

*(5) AI Testing.* We execute the test suite generated by the agent using `pytest`. We verify both the presence and executability of the generated tests. A violation is recorded if tests are missing, cannot be executed, or fail. We also measure the backend statement coverage achieved by the AI-generated tests.

*(6) Project Structure*. We manually inspect the repository structure, checking if the required top-level directories (`backend/`,`frontend/`, and `tests/`) have been created and that files are placed coherently. This is the only step that requires qualitative human judgment.

Violations found by all the steps of the protocol are recorded as errors, which are assigned to one functional area depending on the cause of the violation: API/backend, persistence, frontend, tests, project structure, or other. If one or more errors are found, we provide the agentic IDE with a structured corrective prompt listing all observed errors and the expected behavior. The generated project is then validated again using the same protocol. This validation-correction loop continues until the implementation satisfies all requirements or the maximum number of corrective iterations is reached. In our experiments, all projects converged before the cap of 10 corrective iterations, with average corrective iterations between 1.13 and 1.73 across IDEs.

The frontend and backend of the application that is finally generated by the Agentic IDEs are also assessed with static analysis tools. For the *Frontend Assessment*, we run Lighthouse [5] to assess on a 0–100 scale the performance, accessibility, best practices, and the Search Engine Optimization (SEO) of the frontend. For the *Backend Assessment*, we run SonarQube [7] using its default configuration to detect reliability, maintainability, security, duplication, and security-hotspot issues on the backend. The metrics are computed as the number of issues detected for each category, with the exception of code duplication, which is measured as the percentage of duplicated code.

*Metrics.* We discuss below how the metrics collected with the validation protocol are used to answer each research question.

**RQ1 (effectiveness).** To measure the effectiveness of the Agentic IDEs, we use three main classes of metrics.

*Errors by functional area:* We record the number of errors detected by applying the validation protocol described in the previous section. We distinguish among functional areas (frontend, backend, persistence, tests, project structure, and other) to reveal specific weaknesses of agentic IDEs. This metric provides a comprehensive assessment of the capability of an Agentic IDE to generate code that satisfies the given requirements.

*Frontend and Backend Auditing:* To study the quality of the frontend and backend, we rely on the output generated by Lighthouse [5] and SonarQube [7], as described in the validation protocol.

**RQ2 (iterations)** To capture the need and effectiveness of the corrective iterations computed by the Agentic IDEs, we compute both the number of iterations completed to generate an actual implementation and the number of regressions incidentally introduced in the code.

*Number of corrective iterations:* We count the number of correction prompts required to reach full compliance, which captures the capability of the agents to self-correct problems.

*Regressions:* We count the number of regressions, that is, the number of times a requirement that was satisfied in a iteration is violated in the follow-up iteration, capturing instability and inconsistency in iterative repair.

**RQ3 (Efficiency)** To measure the efficiency of the Agentic IDEs, we collect two main classes of metrics capturing both the machine time and the human effort spent on the applications.

*Agent execution time (minutes):* We measured the time during which the agent is actively generating or modifying code (excluding idle time), used as a proxy for generation productivity.

*Manual verification time (minutes):* We logged the amount of manual effort spent in completing the actions presented in the Validation Protocol, as a proxy for residual human effort necessary to complete the project implementation through interactions with the AI.

**RQ4 (Evolution)** To consider the impact of model evolution, we repeated the generation of the Note Manager application with an evolved version of the underlying LLM. We then computed all the metrics to assess the generated applications.

| Metric | Application | Github Copilot | Cursor | Windsurf |
|---|---|---|---|---|
| Average Number of Errors per Project | Note Manager | **2,00** | 6,33 | 3,67 |
| | Authentication Module | **0,33** | 1,67 | 1,33 |
| | File Hosting | 1,33 | **1,00** | 2,33 |
| | Real-time Chat | 1,67 | **0,67** | **0,67** |
| | Task Queue | 3,00 | 4,67 | **1,67** |
| | **average** | **1,67** | 2,87 | 1,93 |
| Frontend Accessibility | Note Manager | 90,67 | 88 | 91 |
| | Authentication Module | 88 | 88 | 89,33 |
| | File Hosting | 84,67 | 81 | 79,67 |
| | Real-time Chat | 87 | 90,67 | 90,67 |
| | Task Queue | 91,67 | 91,67 | 88 |
| | **average** | **88,4** | 87,87 | 87,73 |
| Backend Reliability Issues | Note Manager | **0** | 1,33 | 0,66 |
| | Authentication Module | 3,33 | 3 | **1** |
| | File Hosting | **0,66** | 1,33 | **0,66** |
| | Real-time Chat | 3 | 2,33 | **0** |
| | Task Queue | 9 | 11,66 | **4,33** |
| | **average** | 3,2 | 3,93 | **1,33** |
| Backend Maintainability Issues | Note Manager | **1,33** | 2,33 | 3 |
| | Authentication Module | 4,66 | 3 | **2** |
| | File Hosting | **1** | 1,66 | 1,66 |
| | Real-time Chat | 4 | **3,66** | 4,66 |
| | Task Queue | 11 | 13,66 | **6,66** |
| | **average** | 4,4 | 4,86 | **3,6** |
| Backend Coverage | Note Manager | **63,5%** | 51,1% | 48,1% |
| | Authentication Module | 47,3% | **91,7%** | 51,1% |
| | File Hosting | **90,3%** | 88,3% | 83,8% |
| | Real-time Chat | 89,9% | 89,2% | **91%** |
| | Task Queue | 65,1% | **68,1%** | 60,7% |
| | **average** | 71,2% | **77,7%** | 66,94 |

**Table 1: Effectiveness metrics**

# 4 Results

## 4.1 RQ1 - Effectiveness

Table 1 summarizes the effectiveness metrics measured for the three considered Agentic IDEs (reported in the columns) and the five considered applications (reported as rows and repeated for each metric). Row *average* indicates the average value of the given metric across the five projects. Bold values indicate the best metric value obtained for each application. Metric *Average Number of Errors per Project* reports the average number of errors per project generation identified with the validation protocol described in Section 3.4 (fewer is better). Metric *Frontend Accessibility* measures the adequacy of the front end with a value between 0 and 100 (the higher the better). Metrics *Backend Reliability Issues* and *Backend Maintainability Issues* indicate the number of reliability and maintainability issues discovered by SonarQube per project (the lower the better). Metric *Backend Coverage* measures statement coverage of the backend code as obtained with the generated tests.

We reported only the accessibility score of the frontend, omitting the frontend performance, best practices and SEO scores computed by Lighthouse since these three other metrics scored identically for the three agentic IDEs, with all values between 90 and 100, indicating a strong capability of dealing with these frontend concerns, at least to the extent they are captured by Lighthouse. Similarly, for the backend, we have not reported the issues concerning security, duplications, and security hotspots since none of the tools generated any of these issues, as detected by SonarQube. It is surprising to observe how consistently robust all the Agentic IDEs have been with respect to certain aspects.

In terms of the average performance of the agentic IDEs, we could notice that different tools excel in different areas. Copilot generates fewer errors that require human intervention (1,67 in average, against 1,93 of Windsurf and 2,87 of Cursor), and produces a slightly more accessible frontend than others (although differences are minimal for the frontend). Cursor demonstrated to be stronger in test case generation than other tools, achieving an average statement coverage of 77,7%. Finally, Windsurf demonstrated to be stronger in the generation of backend code, producing fewer reliability and maintainability issues than the other tools. These complementarities suggest that multi-agent solutions that combine these agents could be explored to leverage the strengths of each.

The per-project results however show that the superiority of one agentic IDE over the others is not confirmed on a per-project basis, and different agents perform better on specific projects. In particular, Copilot tends to perform better on applications rooted in well-established domains and built upon standard architectural patterns, as the Note Manager, Authentication, and File Hosting applications. Conversely, Cursor and Windsurf alternate as the best performing agentic systems with the application targeting specific domains and built upon distributed architectures, as Real-time Chat and Task Queue are.

In our experiments, the hypothesis that specific domains, which might be less represented in training data, and distributed architectures, which are more difficult to engineer, represent challenging cases, is confirmed across all Agentic IDEs that tend to generate code with a higher number of issues for Real-time chat and even more for Task Queue.

Among the considered metrics, the frontend is always of high quality according to the auditing process. Few recurring frontend issues include accessibility violations (e.g., insufficient color contrast, missing form labels) and minor best-practice violations such as console errors and missing metadata. However, we observed a recurring tendency among the agents to rely on similar interface templates, even when the generated Web applications target completely different domains. The chosen color palettes also show limited variability, often converging toward blue-violet tones. This suggests that, unless explicitly instructed otherwise, agents may exhibit a bias toward specific frontend styles.

Backend and maintainability are good, as long as the domain and the architecture are not too challenging. The issues detected are mostly related to maintainability and reliability (e.g., code smells and potential stability problems), while no critical security vulnerabilities are observed. Test coverage is good although it can be improved, both in average and in specific cases. In particular, test-related errors account for a large fraction of the observed issues, and are concentrated on complex I/O operations, asynchronous behaviors, and edge cases, which are not consistently handled by the generated test suites.

We also investigated for each agentic IDE how the errors detected by the validation procedure are distributed across the internal layers and components of the applications, to understand which areas require more research to be addressed properly. Results are summarized in Table 2.

The persistence layer and the project structure are always generated consistently with the specification. Instead, the backend API, the frontend, and especially the generated tests are the most problematic areas of the application. The API layer often fails to fully comply with the specified requirements, particularly in terms of endpoint behavior and correct handling of requests, especially in more complex scenarios involving asynchronous logic and state management. Agentic IDEs sometimes also struggled to deal with the Python environment. The frontend often faces accessibility issues, such as insufficient color contrast and missing form labels. The tests often fail to cover all required functionalities, particularly edge cases and more complex scenarios such as asynchronous behavior or I/O operations. We observed that Agentic IDEs sometimes quit the goal of producing high-quality tests after a few unsuccessful attempts to generate or execute them. In some cases, they even misinterpreted test results, for example, treating the presence of failing tests as acceptable within the test suite. In practice, in many cases, tests are incomplete or require manual adjustments, confirming that test generation is the most challenging aspect for agentic IDEs.

We finally noticed Agentic IDEs tend to document both their activity and the generated artifacts using markdown files, even if this is not required.

**Answer to RQ1** In our experiments, agentic IDEs have all been successful in generating full-stack web applications, although with some specificities: Copilot performed fewer errors that require human intervention, Cursor excelled in test generation, and Windsurf produced more maintainable and reliable backend code. We also observed that the domain matters: agentic IDEs performed better with well-known and architecturally simpler domains. The generated tests are often problematic, with some issues affecting the backend API and, to a minor extent, the frontend.

## 4.2 RQ2 - Iterations

Table 3 summarizes the behavior of the Agentic IDEs for each target project, showing the number of iterations completed until obtaining a system that passes the validation stage, column *number of corrective iterations*, and the number of regressions introduced while iterating, column *regressions*.

Results show that some corrective iterations are typically required for every project that is generated. In particular, the average number of iterations per project ranged between 0.33 and 3.33. Interestingly, in our experiments, an application that passes the validation protocol could always be obtained with a small number of iterations. This result suggests that the first version of the application that is generated is typically already of high quality, and the introduced inadequacies could be quickly fixed, typically with a single additional iteration, sporadically requiring some extra iterations. Among the assessed Agentic IDEs, Copilot required slightly fewer iterations than Cursor and Windsurf to produce an adequate implementation.

| Experiment | API | Persist. | Frontend | Tests | Struct. | Other | Total |
|---|---|---|---|---|---|---|---|
| Github Copilot | 0.60 | 0.00 | 0.33 | 0.73 | 0.00 | 0.00 | 1.67 |
| Cursor | 1.00 | 0.00 | 0.33 | 1.53 | 0.00 | 0.00 | 2.87 |
| Windsurf | 0.40 | 0.00 | 0.47 | 0.93 | 0.00 | 0.13 | 1.93 |
| **Average** | **0,67** | **0** | **0,38** | **1,06** | **0** | **0,04** | **2,16** |

**Table 2: Errors per project by functional area, reported for each Agentic IDE.**

| Metric | Application | Github Copilot | Cursor | Windsurf |
|---|---|---|---|---|
| Number of corrective iterations | Note Manager | 1.67 | 3.33 | 3.33 |
| | Authentication Module | 0.33 | 1.33 | 1.33 |
| | File Hosting | 1.00 | 1.00 | 1.67 |
| | Real-time Chat | 1.00 | 0.67 | 0.67 |
| | Task Queue | 1.67 | 2.33 | 1.33 |
| | **average** | **1,13** | 1,73 | 1,67 |
| Regressions | Note Manager | 2 | 2 | 4 |
| | Authentication Module | 0 | 0 | 0 |
| | File Hosting | 0 | 0 | 0 |
| | Real-time Chat | 0 | 0 | 0 |
| | Task Queue | 2 | 2 | 0 |
| | **average** | **0,8** | **0,8** | **0,8** |

**Table 3: Self-correction capabilities.**

Not all iterations impose the same effort on the developer. When an issue results in an explicitly logged error, the developer can simply forward the error message to the agentic IDE and request a fix. In contrast, when the issue manifests as a missing reaction, such as the application not responding when a button is clicked, the developer must invest additional effort to gather contextual information about the application's behavior (e.g., backend execution errors) before providing a meaningful problem description to the agent.

It is interesting to notice how regressions are seldom introduced; that is, agentic IDEs were typically able to fix issues without introducing new ones. In some exceptional cases, some regressions were introduced. For instance, regressions typically manifest as requirements that were previously satisfied becoming violated in later iterations. Common examples include breaking API endpoints that initially worked, introducing inconsistencies in frontend behavior after UI fixes, or degrading test suites, for example, tests that previously passed becoming failing or incomplete. These regressions often stem from corrective changes that propagate across multiple components, particularly in complex scenarios involving asynchronous logic or multi-component interactions, where addressing one issue may inadvertently affect others.

**Answer to RQ2** Iteratively improving the generated code is necessary to obtain valid implementations, although only few iterations are typically needed by Agentic IDEs. Iterative improvement rarely introduces regressions.

## 4.3 RQ3 - Efficiency

Table 4 shows the *agent execution time*, which is the time needed by each Agentic IDE to generate the application, and the *manual verification time*, which is the amount of time needed to manually inspect the output according to the validation protocol.

| Metric | Application | Github Copilot | Cursor | Windsurf |
|---|---|---|---|---|
| Agent execution time (mm:ss) | Note Manager | 17:38 | 05:20 | 17:39 |
| | Authentication Module | 05:16 | 05:05 | 12:59 |
| | File Hosting | 07:52 | 02:20 | 13:02 |
| | Real-time Chat | 08:12 | 05:44 | 06:23 |
| | Task Queue | 09:38 | 14:15 | 20:31 |
| | **average** | 9:43 | **6:33** | 14:07 |
| Manual verification time (mm:ss) | Note Manager | 14:00 | 19:05 | 15:37 |
| | Authentication Module | 08:16 | 08:55 | 11:22 |
| | File Hosting | 08:27 | 05:32 | 11:59 |
| | Real-time Chat | 11:37 | 05:44 | 08:57 |
| | Task Queue | 19:29 | 19:02 | 14:04 |
| | **average** | 12:22 | **11:40** | 12:24 |

**Table 4: Efficiency.**

Cursor has consistently been the fastest tool, with an average agentic execution time of 6 minutes and 33 seconds. Copilot is the second fastest tool, with an average execution time of 9 minutes and 43 seconds (×1,48 with respect to Cursor). Windsurf has been the slowest, with an average execution time of 14 minutes and 7 seconds (×2,16 with respect to Cursor).

The absolute execution time is always small, considered a working implementation is obtained without writing a single line of code: 2 minutes and 20 seconds in the best case (Cursor with the file hosting application), and 20 minutes and 31 seconds in the worst case (Windsurf with the Task Queue application). If Agentic IDEs are systematically employed to generate code, the observed relative differences may become meaningful, with, for instance, Cursor potentially generating nearly double the amount of code generated by Windsurf.

Although their overall performance has been efficient, we observed that Agentic IDEs tend to rely heavily on Linux command-line conventions when interacting with the codebase. This bias can introduce friction or slowdowns when the agent operates in environments that differ from its assumed setup.

| Metric | Application | Github Copilot | Cursor | Windsurf |
|---|---|---|---|---|
| Errors | Note Manager | 2,00 | 6,33 | 3,67 |
| | Note Manager 2 | 0,00 | 2,00 | 0,00 |
| | **delta** | **+100%** | **+68%** | **+100%** |
| Frontend Accessibility | Note Manager | 95 | 94,25 | 94,25 |
| | Note Manager 2 | 88 | 88 | 93,33 |
| | **delta** | **-7%** | **-7%** | **-1%** |
| Backend Reliability | Note Manager | 0 | 1,33 | 0,66 |
| | Note Manager 2 | 0,66 | 0,66 | 0 |
| | **delta** | **-100%** | **+50%** | **+100%** |
| Backend Maintainability | Note Manager | 1,33 | 2,33 | 3 |
| | Note Manager 2 | 2 | 2,33 | 2 |
| | **delta** | **-50%** | **0%** | **+33%** |
| Backend Coverage (%) | Note Manager | 63,46 | 51,06 | 48,1 |
| | Note Manager 2 | 96,86 | 95,8 | 99,63 |
| | **delta** | **+53%** | **+88%** | **+107%** |
| Number of corrective iterations | Note Manager | 1,67 | 3,33 | 3,33 |
| | Note Manager 2 | 0 | 0,33 | 0 |
| | **delta** | **+100%** | **+90%** | **+100%** |
| Regressions | Note Manager | 2 | 2 | 4 |
| | Note Manager 2 | 0 | 0 | 0 |
| | **delta** | **+100%** | **+100%** | **+100%** |
| Agent execution time (minutes) | Note Manager | 17:38 | 05:20 | 17:39 |
| | Note Manager 2 | 02:44 | 03:36 | 02:56 |
| | **delta** | **+84%** | **+33%** | **+83%** |
| Manual verification time (minutes) | Note Manager | 14:00 | 19:05 | 15:37 |
| | Note Manager 2 | 03:51 | 05:50 | 04:30 |
| | **delta** | **+73%** | **+69%** | **+71%** |

**Table 5: Evolution**

Manual verification time is similar between Agentic IDEs. Cursor required slightly less effort in average than competing approaches: 11 minutes and 40 seconds, against 12 minutes and 22 seconds of Copilot and 12 minutes and 24 second of Windsurf.

It is important here to remark that, although the user is not writing code, the careful application of the validation protocol that requires the inspection of all the layers in the generated code according to multiple perspectives requires non trivial effort. This is an important cost factor that must be considered when assessing agentic solutions. In fact, the obtained code is not granted to be correct, and in practice, as observed in our experiments, it is not. Thus developers have to check the code and iterate code generation a few times, providing detailed instructions about the aspects that must be fixed.

**Answer to RQ3** Although an application could always be obtained in less than 20 minutes, the efficiency of agentic IDEs may vary significantly (up to a ×2 factor). Significant manual effort is necessary to inspect the generated code and provide instructions about the necessary improvements that must be implemented.

## 4.4 RQ4 - Evolution

RQ4 examines how much effectiveness depends on improvements in the underlying language models over time. The experiments span two model generations, and the results show substantial progress on consolidated patterns, with remaining limitations that appear structurally persistent.

A key contribution to RQ4 is the repetition of the Note Manager application development where functional requirements are held constant while the tool and model ecosystem evolves from Claude Sonnet 3.5 to Claud Sonnet 4.5. In Table 5, the evaluation metrics of the two applications' versions are reported together with the variation delta. The repeated design yields a large reduction in errors and in required corrections. Agent execution times also decline sharply for two IDEs (84–83% reduction), and corrective iterations are almost eliminated (0.00–0.33 corrections on Note Manager 2, versus 1.67–3.33 in the original Note Manager). Regressions, which were prominent in the earliest experiment (8 total), are fully eliminated in the repeated Note Manager run.

In addition to effectiveness and efficiency gains, the evolution of the underlying models also leads to a substantial improvement in test quality. Test coverage increases consistently across all IDEs (from 48–63% to 95–99%), indicating that more recent models are significantly more capable of generating comprehensive and executable test suites. This result partially mitigates one of the main limitations observed in earlier experiments, where test generation was the dominant source of errors.

However, improvements are not uniform across all quality dimensions. Backend code quality metrics such as reliability and maintainability show more heterogeneous trends, with some IDEs exhibiting marginal improvements and others slight degradations. Similarly, frontend accessibility scores slightly decrease across all IDEs, suggesting that gains in functional correctness and completeness may come at the expense of non-functional aspects such as usability and accessibility.

Overall, these results suggest that model evolution has a relevant impact on the effectiveness of agentic IDEs, potentially enabling near-perfect performance on consolidated application patterns with minimal human intervention. At the same time, certain limitations—particularly related to non-functional quality attributes and complex architectural reasoning—appear to be more persistent and less sensitive to improvements in the underlying models.

**Answer to RQ4** Model evolution is a key factor for the successful generation of software applications from requirements. We observed improvements in nearly all areas when moving from Claude Sonnet 3.5 to Claude Sonnet 4.5, with some little degradation on specific indicators.

## 4.5 Threats to Validity

*Construct validity.* A first threat concerns whether the metrics operationalize the intended constructs (e.g., effectiveness, efficiency, and self-correction) without systematic bias. Error counts are derived from requirement-level checklists and are therefore sensitive to how requirements are specified, interpreted, and mapped to error categories (API, tests, frontend, persistence, etc.). Although the checklists aim for operational clarity, borderline cases (e.g., partial satisfaction, performance optimizations, or alternative acceptable designs) can introduce classification noise. Similarly, interaction time incorporates human validation and corrective prompting, which may vary with evaluator experience and with the clarity of observable failures. This threat is mitigated by the use of detailed validation checklists with explicit acceptance criteria, applied uniformly across all experiments, and by complementing manual assessment with automated tools (Lighthouse and SonarQube) that provide standardized measurements.

*Internal validity.* The experimental timeline introduces potential confounding factors, as model versions evolve, IDE implementations change, and the set of evaluated IDEs is not constant across all experiments. As a result, observed differences may conflate model evolution, IDE-level improvements, and task characteristics. In addition, LLM-based generation is inherently stochastic, and residual variance may remain despite repeated runs. This threat is mitigated through a controlled design with standardized prompts, technology stack, validation protocol, and measurement procedures, combined with three repetitions per configuration. Moreover, RQ1-RQ3 experiments are executed using the same model. The repeated Note Manager in RQ4 experiment further isolates model evolution by keeping requirements constant.

*External validity.* Generalizability is limited by the scope of the evaluation tasks and by the chosen technology stack. The selected application scenarios consist of five full-stack Web applications that cannot be representative of other classes of systems (e.g., large-scale industrial software systems). They cover common Web-development layers, including backend APIs, frontend interfaces, persistence, and automated tests, and progressively introduce concerns such as authentication, file handling, real-time communication, and asynchronous background processing. However, they do not capture several characteristics of industrial systems, such as large monorepositories, long-lived codebases, microservice architectures, production deployment pipelines, organizational coding conventions, complex third-party integrations, observability requirements, or domain-specific constraints. Therefore, our findings should be interpreted as evidence about controlled end-to-end application-generation tasks, not as evidence that agentic IDEs can generate production-ready industrial systems. The stack based on FastAPI and plain HTML/JavaScript was selected to balance realism, reproducibility, and experimental control. This choice limits generalizability to framework-heavy stacks such as React, Spring Boot, or Node.js. Results may also differ under other model families or deployment modalities. We partially mitigate these threats by studying recurring Web-development scenarios of increasing architectural and operational complexity, and by releasing the experimental material to support replication. Further experiments are needed on larger systems, alternative frameworks, and different domains to generalize our observations.

*Conclusion validity.* While the dataset size and the descriptive nature of the analysis limit the strength of claims, the results highlight consistent qualitative patterns, such as persistence being systematically correct, test generation being the main source of errors, and distributed architectures being challenging. Time measurements may also be affected by environmental factors, and regression counting depends on consistent tracking across iterations. This threat is mitigated by relying on multiple complementary metrics (effectiveness, efficiency, and self-correction), providing converging evidence across different perspectives and focusing on robust trends.

## 5 Actionable Results

The results presented in this paper generates some actionable results that are summarized below:

- **Development teams may consider starting** their activity from full-stack web applications fully generated by Agentic IDEs: The effectiveness and efficiency of Agentic IDEs are changing software engineering practices. Indeed, it is now an appealing option to start with an automatically generated application rather than developing one from scratch. Our results reveal that the generated code is not only a possible starting point for development, but also already satisfies several quality aspects (e.g., maintainability or reliability) that make it potentially eligible for professional development.
- **The generation of automated tests is an open challenge**: Although relatively simple code is generated, in our experiments agentic IDEs consistently fail to achieve high coverage, with all tools performing poorly on the Task Queue application, which is the application addressing a specific domain with the most complex architecture. The generation of thorough test suites able to deeply exercise the code is essential to validate automatically generated code and increase developers' confidence on the correctness of the results. More research is needed to improve the self-validation capabilities of agentic IDEs.
- **Domain-specific and architecturally-complex applications challenge agentic IDEs:** While agentic IDEs performed generally well in our experiments, the results with the Real-time chat, and especially with the Task Queue, have been worse than others according to multiple perspectives (e.g., issues to be detected and fixed, code maintainability and reliability, tests). This suggests that more effective approaches might be needed to handle domains and technical contexts that are less common.
- **Self-corrective mechanisms are necessary:** Our experiments show that multiple iterations were almost always necessary to obtain a satisfactory implementation. On one hand, this result shows that the code generation process cannot be designed as a sequential process without assessment and iteration. On the other hand, the few iterations that were necessary to complete the generation show that self-correction policies often converge towards a stable implementation. This motivates research on ever more sophisticated assessment and correction mechanisms that can be designed to address the qualitative aspects that must be enforced on the implementation.
- **Applications can be implemented without coding:** In our study, we obtained applications without writing code, yet we had to inspect the generated code and prompt for some corrective actions. However, these results confirm that approaches like vibe coding [23], where developers do not write code but uniquely orchestrate the agentic IDE with natural language requests, are a promising direction for future research.
- **Reducing the required human effort is a priority:** In our experiments, we measured both the amount of time spent by the AI in performing code-related tasks and the human time spent in validating the code. Based on our results, human effort is as much as, and often higher than, AI time. We speculate that the more the requirements to be satisfied are specific, the higher the human effort would be, and thus our results represent an under-approximation of the human effort that might be needed when interacting with AI to obtain working implementations. Overall, this suggests that improvement to the code generation process should prioritize the reduction and optimization of human effort, rather than increasing AI efficiency.
- **Model evolution is still a key factor:** Research on better AI models is still extremely relevant and useful, as confirmed by our comparative study with two different generations of Claude Sonnet models.
- **The choice of the Agentic IDE is a relatively important factor:** Our study reports differences along all the considered dimensions for the three experimented agentic IDEs. However, the observed differences were small, suggesting that the choice of the agentic IDE is a relatively important factor.
- **Agentic IDEs have implementation biases:** If not instructed carefully, Agentic IDEs may exhibit systematic biases. For example, we observed a shared tendency toward specific frontend styles and a recurrent use of Linux command-line instructions during code generation.

# 6 Takeaways

Our study results in a set of takeaways for practitioners and researchers.

**Takeaways for Practitioners** Agentic IDEs perform best when requirements align with established Web-development patterns such as CRUD operations, authentication flows, and conventional request–response APIs. In these cases, the generated applications are close to the expected behavior and typically require only limited corrective interaction. They are therefore *well-suited for rapid prototyping, scaffolding, and self-contained modules*. On the contrary, tasks involving real-time communication, background workers, retry logic, cancellation, asynchronous execution, or complex state transitions remain challenging. The agent may still produce plausible and executable code, but correctness must be carefully validated.

Although agents often produce executable test suites that may support the validation of the generated code, the *AI-generated tests might be incomplete (e.g., requirements not exercised by tests) and imprecise (e.g., wrong oracle)*. Developers should thus run the generated tests but also inspect them critically and complement them with independent assessment activities (e.g., user-implemented tests, automatic static and dynamic analysis, manual code inspection).

Agentic IDEs can exhibit systematic biases that lead to the generation of overly similar code artifacts, such as similar front-end structures, when not instructed with sufficient precision. This homogenization effect *may result in different organizations unintentionally producing applications with comparable look-and-feel*, which is undesirable from a practitioner's perspective. *Designing prompts that explicitly guide and personalize* the generation process is, therefore, essential to mitigate these biases and foster more diverse and context-appropriate outputs.

In a nutshell, even when agentic IDEs generate complete applications quickly, systematic developers' validation remains essential. Agentic IDEs shift effort rather than eliminate it: *less time writing boilerplate, more time verifying requirements, assessing tests, debugging, and supervising architecture*.

**Takeaways for Researchers** We observed that Agentic IDEs sometimes struggle with architectures involving asynchronous logic, background processes, or multi-component coordination. Enhancing the capability of tools to *reason on multiple components simultaneously, maintaining consistency of changes* may expand the applicability of Agentic IDEs.

Improving Agentic IDEs requires not only better code generation but also more reliable test-oracle construction, edge-case identification, and coverage assessment. Tools should help ensure that *generated tests are adequate*, actually validating the intended requirements, and correct.

Since validation is a core part of supervised agentic development, tools should incorporate *mechanisms that guide developers through systematic checking of generated artifacts*, reducing the risk of misplaced trust. Instead, the definition and actuation of a validation protocol for the code generated by Agentic IDEs is currently entirely dependent on the development team.

Moreover, *understanding and quantifying the validation burden* is essential for realistic productivity claims. Research should explore how validation effort interacts with agentic generation, corrective prompting, debugging, and architectural supervision.

# 7 Conclusions

This paper presented a controlled empirical evaluation of agentic IDEs for the generation of full-stack web applications. Across 48 generated projects spanning five application scenarios and three agentic IDEs, we assessed effectiveness, efficiency, and self-correction capabilities using a combination of functional validation, automated analysis, and manual inspection.

Results show that agentic IDEs have reached high maturity on consolidated architectural patterns (e.g., CRUD and authentication), where they can generate nearly complete and correct applications with minimal corrective effort. In these settings, errors are few, regressions are rare, and generation time is low, confirming that agentic workflows can significantly accelerate development.

However, two structural limitations persist. First, automated test generation remains a significant source of errors, accounting for nearly half of all observed issues. While newer models improve coverage, they still struggle with edge cases, asynchronous behaviors, and I/O-intensive scenarios. Second, distributed and coordination-intensive architectures remain challenging: tasks involving concurrency, state management, and asynchronous workflows exhibit higher error rates and occasional regressions, indicating limits in current agentic reasoning.

The comparative analysis highlights that no IDE is universally superior. Instead, tools exhibit complementary trade-offs: some prioritize speed, others stability or code quality. At the same time, the assessment of the impact of model evolution shows that improvements are largely driven by advances in the underlying LLMs.

Overall, agentic IDEs are effective for rapidly generating initial implementations and standard components, but they do not eliminate the need for developer supervision. Rather, they shift effort toward prompt engineering, validation, testing, and architectural reasoning, especially as application complexity increases.

Future work concerns with considering extending the study to other domains, to generalize findings beyond the scope of full stack Web applications. Moreover, additional scenarios deserve to be validated. In particular, requirements can be stated at different level of granularity and understanding the capability of agentic IDEs to follow fine-grained specification is another relevant challenge.

## Data Availability Statement

The tools, scripts and empirical data necessary to repeat the results reported in this paper are available at https://zenodo.org/records/19208143